# Domains in Three-dimensional Ferroelectric Nanostructures: Theory and Experiment

G. Catalan,* J. F. Scott,* A. Schilling,[+] J. M. Gregg[+]

* Department of Earth Sciences, University of Cambridge,
Cambridge CB2 3EQ, United Kingdom

[+]Centre for Nanostructured Media, School of Mathematics and Physics, Queen's University Belfast, Belfast BT7 1NN, Northern Ireland, United Kingdom

## Abstract

Ferroelectric random access memory cells (FeRAMs) have reached 450 x 400 nm production (0.18 µm$^2$) at Samsung with lead zirconate-titanate (PZT), 0.13 µm$^2$ at Matsushita with strontium bismuth tantalate (SBT), and comparable sizes at Fujitsu with $BiFeO_3$. However, in order to increase storage density, the industry roadmap requires by 2010 that such planar devices be replaced with three-dimensional structures. Unfortunately, little is known yet about even such basic questions as the domain size scaling of 3-d nanodevices, as opposed to 2-d thin films. Here we report the experimental measurement of nano-domains in ferroelectric nanocolumns, together with a theory of domain scaling in 3-d structures which explains the observations.

In the continuing race to decrease the size of electronic devices, the development of thin film technologies and the achievement of ever-decreasing thicknesses have played a central role. Ferroelectrics have not escaped this trend, as the reduction of film thickness has obvious advantages for practical applications in terms of increasing capacitance for capacitor and DRAM applications, and decreasing coercive voltages for NVFRAM applications. However, it seems that, at least at the cutting-edge laboratory level, the ultimate limits of ferroelectric film thickness reduction have already been reached for planar devices. Films as thin as 3 unit cells have been shown to be ferroelectric [1], and ferroelectric/paraelectric superlattices with a stacking periodicity of 1 unit cell are also ferroelectric [2]. And, while the dielectric constant generally decreases for thin films in the sub-micron region, this decrease is now considered to be mostly extrinsic, with single crystal films as thin as 70nm displaying perfectly bulk-like properties [3].

The next step in device miniaturization is thus not in the reduction of film thickness, but in the achievement of ever-smaller 3-d structures. The technology roadmap in the capacitor industry actually indicates that 2010 should be the year in which 3-d capacitor structures such as trenches or nanotubes should take over planar structures [4]. Wouters' group at IMEC has fabricated commercial capacitors in which both the ferroelectric layer and the top electrode are bent over to produce a 3-d mushroom-shaped structure [5]; this gives an increase of 50% in the capacitance per square micron of chip area. Similarly, the group of Ishiwara and Funakubo, in collaboration with Samsung, have produced 3-d trenched capacitor electrodes for DRAMs [6]. Research into such types of structures is still, however, incipient, with relatively few works in the last few years addressing these and other three-dimensional morphologies such as ferroelectric nano-rods, nanotubes and nano-columns [5-10]. It is worth also mentioning that some of these, such as the nanotubes, will find application not just in capacitors and memory devices, but also for microfluidics, such as ink-jet print heads and drug-delivery systems [11, 12].

Here we have turned our attention to the problem of the scaling of ferroelectric domain size as a function of the lateral dimensions of the ferroic material. The interest in this problem is more than purely academic: domain morphology is essential to rationalize dielectric and electromechanical properties of ferroelectrics, and the scaling of domain size is correlated to the scaling in other domain-related properties such as coercive fields and switching kinetics [12].

The structures analyzed in this work are tetragonal columns with the sides parallel to the main crystallographic planes of ferroelectric $BaTiO_3$ single crystal. These were cut out from a commercially obtained single crystal using focused ion beam milling and ulterior annealing to get rid of implanted Ga ion impurities [13]. The crystals were cut in a two-stage process: first, thin slabs of different thickness were made, and then several columns with different lateral sizes were carved out of each lamella. An example of one such structure is shown in Figure 1. The detail and a schematic diagram of the domain symmetry are depicted in figure 2. The "front" faces (X-faces in our notation) have domain walls at 45 degrees with respect to the crystallographic planes, typical of tetragonal in-plane 90-degree domains (also called $a_1$-$a_2$ domains). The domain structure is the same as observed in the surfaces of the lamellae [14], and it probably propagates from the lamellae into the column via the attachment points. But although the symmetry is the same as in the lamellae, the domain size is different, and it changes as a function of the lateral dimension "$y$" [15].

The problem of domain scaling was first analysed by Landau and Lifshitz [16] and Kittel [17], who showed that for thin crystals and films the ferromagnetic domain size ($w$) scales as the square root of the crystal thickness ($d$). The proportionality between $w^2$ and $d$ is in fact a universal property of all ferroics [12,14, 16-22], with the scaling factor being directly proportional to the thickness of the domain walls [12, 14, 22, 23]. To rationalize the results for the nanocolumns, however, it is necessary to extend Kittel's ideas to 3-dimensional shapes.

In order to calculate equilibrium domain size, the surface energy and domain wall energy for the entire crystal must be minimized. The surface energy density scales in direct proportion to the domain size $w$. This is true for demagnetization energy in ferromagnets

[17], depolarization energy in ferroelectrics [18] and elastic energy in ferroelastics [19, 20] –and, as a matter of fact, also for non-ferroic materials such as martensitic steels [21]. The surface energy density in the two X-faces is then

$$F^X_{domain} = U_x w,$$

where $U_x$ is a constant proportional to the energy density. In the X-faces, this should only be due to elastic energy, as the in-plane orientation of the polarization means that there is no depolarization. To compute the surface energy of the two X-faces we simply multiply the energy density times the area of the two X-faces:

$$E^X_{domain} = F^X_{domain} \times 2xy = 2U_x wyz.$$

Likewise, in the Y-faces we have $E^Y_{domain} = 2U_y wxz$, where $U_y$ contains both elastic and depolarization contributions.

The energy of the domain walls in the crystal is proportional to their size and to their number. The area of each domain wall is $S = \sqrt{2}xy$ and the number of domain walls is $N = \frac{z}{w}$. Let $\sigma$ be the energy density per unit area of domain wall [23]; the total energy of all the domain walls in the column will then be:

$$E_{walls} = \sigma S N = \sigma \sqrt{2} xy \frac{z}{w}$$

Adding the surface and the domain wall terms, the total energy of the crystal is

$$E = \sigma \sqrt{2} xy \frac{z}{w} + 2zw(U_x y + U_y x)$$

Minimizing this energy with respect to the domain size $w$ leads to the final equilibrium condition:

$$w^2 = \frac{\sqrt{2}}{2} \frac{\sigma}{\frac{U_x}{x} + \frac{U_y}{y}} \tag{1}$$

There are two important limits:

1) The thin film lamellae are described by the limit y>>x. In this case we recover $w^2 = \frac{\sqrt{2}\sigma}{2U_x} x$, which is precisely Kittel's result for thin films.

2) In the columns, the two length scales are similar (x~y). The surface energy, however, is not, as the Y-face has an energy contribution from depolarization. Thus, the relevant limit is $U_y$>>$U_x$, for which eq. 1 becomes $w^2 = \frac{2\sigma}{\sqrt{2}U_y} y$.

This result implies that, in columns where $x$ and $y$ are relatively similar, the lateral size $y$ (the one perpendicular to the face with highest surface energy) will significantly dominate the scaling behavior, as observed [15]. Furthermore, since the surface energy constant $U_y$ is bigger than the energy constant $U_x$, the slope of the Kittel plots for the columns will be smaller than for the films, as also observed.

Clearly, though, these are only extreme cases, and in general the experimental results will be best described by eq. 1. It is desirable to find a way of representing the results in order to produce a plot that is as visually useful as the Kittel plots (straight lines for $w^2$ as a function of thickness), but which can be used for columns instead of films. This can be done by rearranging the terms in eq. (1):

$$w^{-2} = \frac{2}{\sqrt{2\sigma}} \left( U_x x^{-1} + U_y y^{-1} \right) \tag{2}$$

The experimental results for the domain size as a function of lateral dimension have been plotted in Figure 3 according to this format (that is, the inverse of $w^2$ as a function of the inverse of the lateral dimensions). In the figure we have added the least-squares fit of the data to eq. 2.

While there is a considerable experimental scatter in the data, the experimental trends are reasonably well reproduced. Furthermore, the parameters of the surface fit can now be compared with other experimental results. In particular, the gradient of the intercept between the surface and the plane $y^{-1}$=0 of the plot can be compared with that of standard Kittel plots for thin films. In our case, the intercept has a gradient of 0.062nm$^{-1}$, the inverse of which is ~16nm. In the case of BaTiO$_3$ thin films, the proportionality constant between $w^2$ and thickness $d$ was found to be ~10nm [14], but the periodicity in that case was measured perpendicular to the domain walls, whereas in our work we have measured it parallel to the sides of the column and hence at 45 degrees to the domain walls. Thus, a $\sqrt{2}$ correction factor has to be included in the comparison. With this correction, we obtain a proportionality constant of ~11nm for the columns, very close indeed to that measured in the films.

There are some limitations to the applicability of the above calculations. First, we have neglected the surface contribution from the Z-face, $E^z_{domain} = 2U_z wxy$. Adding this leads to the generalized expression for a tetragonal grain of any size:

$$w^2 = \frac{\sqrt{2}}{2} \frac{\sigma}{\frac{U_x}{x} + \frac{U_y}{y} + \frac{U_z}{z}} \tag{3}$$

Since in the pillars $z$ is much larger than $x$ and $y$, the Z-face contribution to eq. 3 can be neglected. For nanocrystals of smaller aspect ratios, however, it must be included.

The second caution is that we have assumed that the domain structure consists of parallel 90° domains. If, for example, the domain structure was not parallel but a zig-zag, such as the column on the left of Figure 1, then the energy constant $U_y$ may be different due to the lower depolarization energy, and thus the optimum domain size would also be different. This difference in the value $U_y$ for parallel and zig-zag configurations probably accounts for at least some of the scatter in the experimental data.

Our model also neglects the energy contributions from the edges and corners of the columns, which are different from the contributions in the middle of the faces. These will become more important as the columns become narrower. The narrowest columns examined were ~100×100nm. Although the experimental results show no obvious departure from eq. 1 down to the smallest size, the scatter in the data is too large to establish subtle changes in the trends; future work focusing specifically on edge effects is clearly desirable.

Finally, it is worth noticing that, without external stress, the most stable domain configuration in a free-standing column should in fact be one in which the polarization lies along the z-direction (long axis), because in this way both the depolarization energy and the domain wall energy are minimized. That this is not the case in our ferroelectric columns is surprising, but it is consistent with previous results for free-standing lamellae, where a mono-domain configuration with in-plain polarization would also have been the minimum energy state, yet $a_1$-$a_2$ domains were seen instead. The most likely explanation is that, at the phase transition, polarization nucleates in different sites of the lamellae, and the domains then grow laterally until they meet each other. Adjacent domains with orthogonal polarization impose elastic stress on each other, providing the driving force for the splitting into smaller twins.

In summary, we have studied the scaling of periodic domain structures in ferroelectric single-crystal pillars as a function of their lateral dimensions. The experimental results for $BaTiO_3$ were rationalized in terms of minimization of the total free energy of the crystal, including the contributions from domains and domain walls. The proposed equations extend the range of applicability of Kittel's law to ferroic domains in 3-d shapes. An important result is that the two in-plane directions are not equivalent in the modeling, with dominance determined by depolarization effects. The combination of the experimental method (FIB slicing of single crystals) and the generalized scaling equations provide a new tool with which to analyze domain scaling in other ferroic 3-d structures.

The work of G.C is funded by the EU under the Marie Curie Fellowship programme. JFS and JMG acknowledge support from the EPSRC and Invest Northern Ireland.

# FIGURES

**Figure 1:** Geometry of BaTiO$_3$ columnar structures cut with FIB. Ferroelectric 90$^o$ domains can be seen inside the columns.

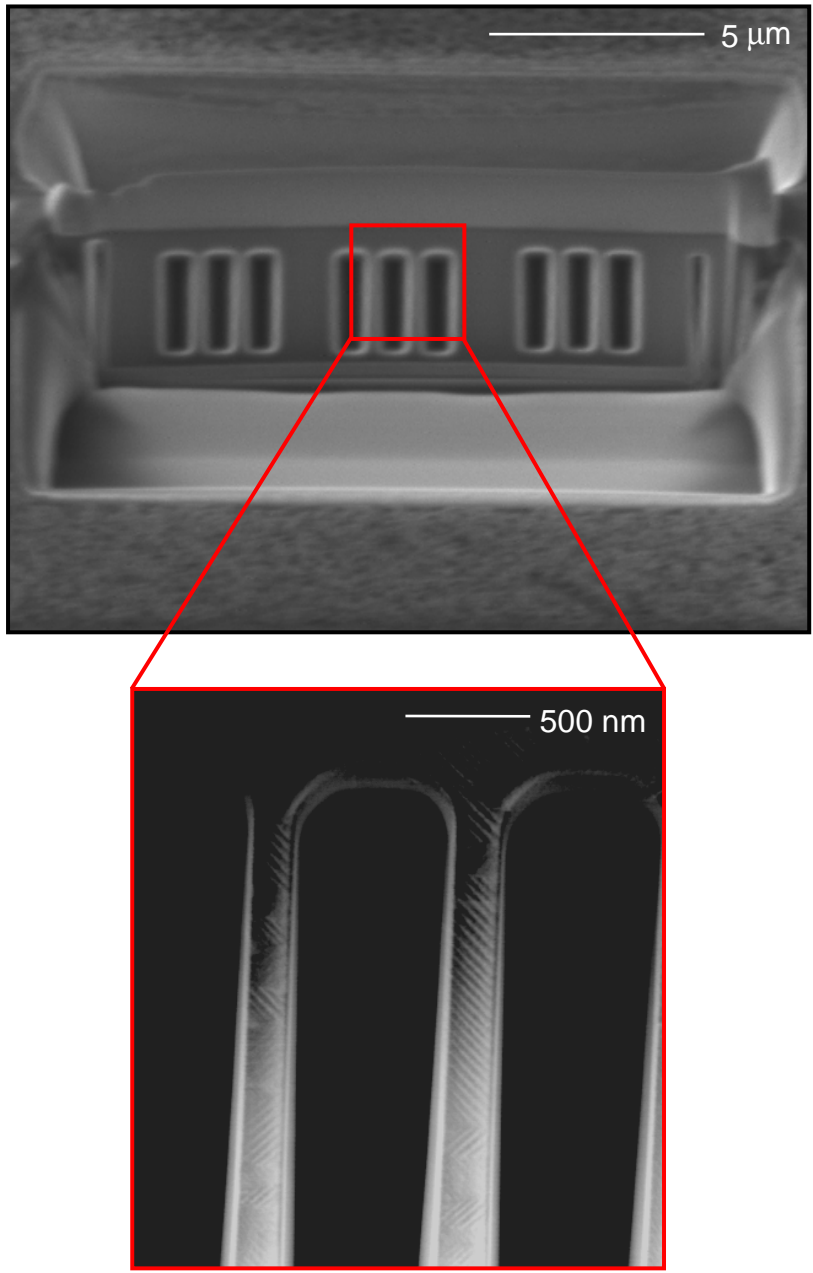

**Figure 2:** Domain configuration in the tetragonal pillars, and schematic of the domain symmetry.

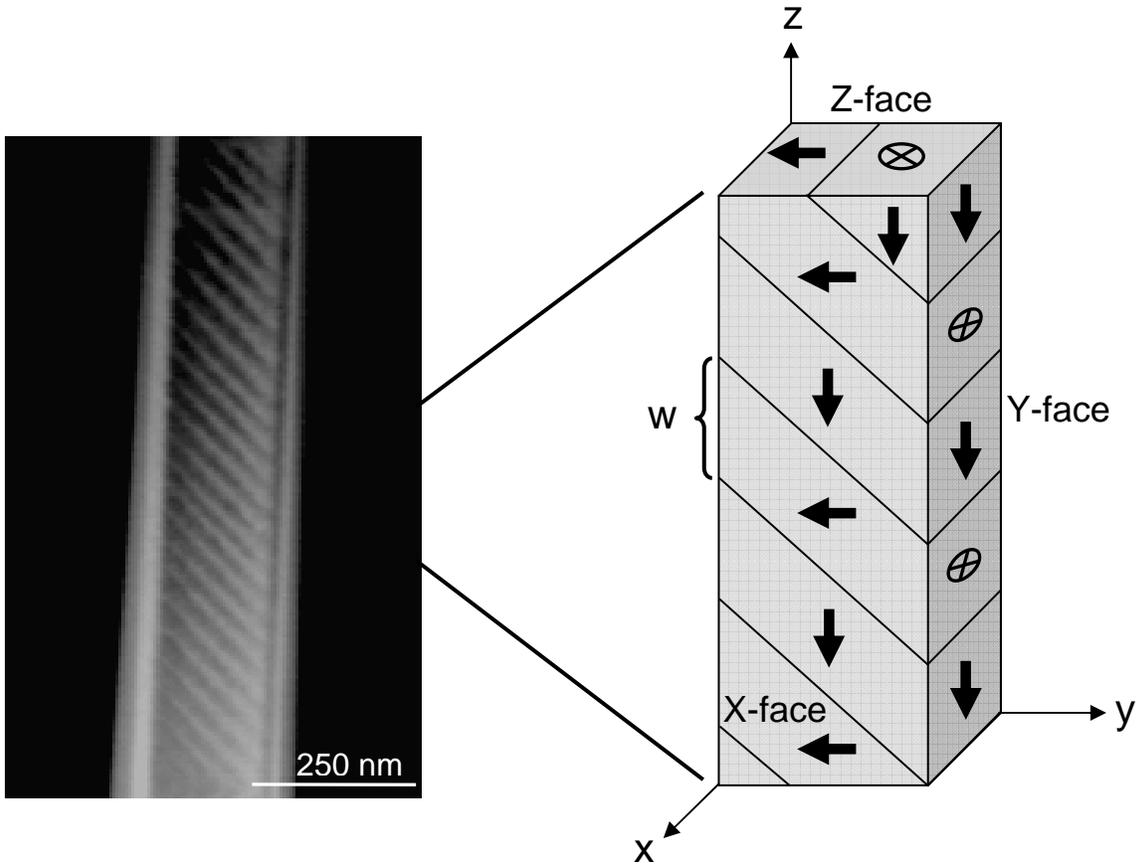

**Figure 3:** Plot of domain size as a function of lateral dimensions for the pillars. The flat surface is the least-squares fit to the 3-d scaling law.

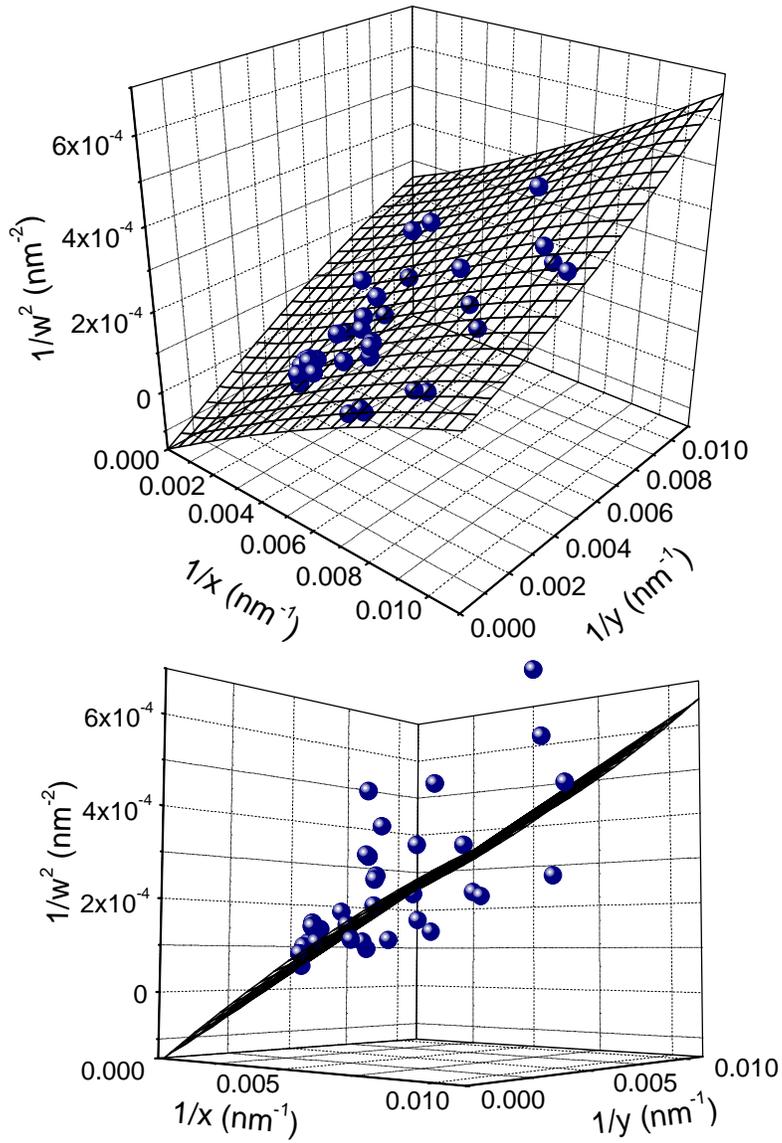